\documentclass[reprint,aps,prb,english,superscriptaddress,nofootinbib]{revtex4-2}

\usepackage{graphicx}
\usepackage{bm,amssymb,amsmath,mathrsfs,amstext,latexsym}

\usepackage{physics}
\usepackage{enumitem}
\usepackage{graphicx}

\usepackage[usenames,dvipsnames]{xcolor}
\usepackage[colorlinks=true,citecolor=blue,urlcolor=blue,linkcolor=blue]{hyperref}
\usepackage[normalem]{ulem}
\usepackage{booktabs}
\usepackage[caption=false]{subfig}
\usepackage{extarrows}
\usepackage{adjustbox}
\usepackage{makecell}

\let\Re\relax
\let\Im\relax
\DeclareMathOperator{\Re}{Re}
\DeclareMathOperator{\Im}{Im}
\let\tilde\relax
\newcommand{\tilde}[1]{\widetilde{#1}}

\begin{document}

\title{Quantum transmission in 1D disordered stealthy hyperuniform \\
Kronig--Penney-like models}

\author{Shaobing Yuan}
\affiliation{Department of Chemistry, Princeton University, Princeton, New Jersey 08544, USA}
\affiliation{Princeton Institute for the Science and Technology of Materials, Princeton University, Princeton, New Jersey 08544, USA}

\author{Carlo Vanoni}
\affiliation{Department of Physics, Princeton University, Princeton, New Jersey, 08544, USA}

\author{Salvatore Torquato}
\affiliation{Department of Chemistry, Princeton University, Princeton, New Jersey 08544, USA}
\affiliation{Princeton Institute for the Science and Technology of Materials, Princeton University, Princeton, New Jersey 08544, USA}
\affiliation{Department of Physics, Princeton University, Princeton, New Jersey, 08544, USA}
\affiliation{Program in Applied and Computational Mathematics, Princeton University, Princeton, New Jersey 08544, USA}



\begin{abstract}
  Disordered stealthy hyperuniform (SHU) systems are emerging as platforms for controlling classical and quantum transport due to vanishing single-scattering contributions in the stealthy regime.
    Here, to our knowledge, we present the first perturbation theory of one-dimensional quantum transport through a continuum of identical point scatterers with SHU positional disorder.
    Our perturbation theory indicates the exact cancellation of all the lower-order terms due to the Fourier-space stealthy condition $S(q)=0$, and the further suppression of the lowest non-vanishing fourth order in the Lyapunov exponent $\lambda(k)$ due to the real-space hyperuniform constraint $\lim\limits_{D\to\infty}\sigma_D^2=c<\infty$.
    Our findings presents universal implications of the general wave-propagation phenomena in disordered SHU systems.
\end{abstract}

\maketitle

\textbf{Introduction.---}
The transport of waves and particles through disordered media is a central problem in condensed matter physics, with implications for electronic conduction, photonic and phononic propagation, cold-atom systems, and heterogeneous materials. A fundamental question is how the statistical structure of disorder controls the crossover between transmission and localization. 
In one-dimensional systems, uncorrelated disorder always leads to Anderson localization~\cite{Anderson1958absence,abrahams1979scaling,lee1985disordered,evers2008anderson}, however weak it might be.
On the other hand, spatial correlations in the disordered potential can induce coherence effects in the scattering processes and thereby modify the usual localization scenario~\cite{Lugan2009Onedimensional,Piraud2012Anderson,morpurgo2025localizationtransitioninteractingquantum,izrailev_localization_1999,izrailev_anomalous_2012,Izrailev1995Hamiltonian}.
Quasi-periodic systems provide a paradigmatic example, exhibiting extended, critical, and localized states depending on the potential strength and on the underlying incommensurate structure~\cite{Aubry1980Annals,Luschen2018,Ganeshan2015Nearest,Biddle2010Predicted,Goncalves2023Critical}.

Among all these correlated disordered systems, stealthy hyperuniform (SHU) ones have been reported to exhibit unusually weak scattering and high finite-sample transmission over continuous wave-number intervals~\cite{kim_effective_2023,kimTheoreticalPredictionEffective2024,Kim_2024_extraordinary,klattTransparencyAndersonLocalization2026,vanoni2025effective}.
Existing analyses establish low-order cancellations but do not determine whether multiple scattering produces a nonzero Lyapunov exponent~\cite{kim_effective_2023,kimTheoreticalPredictionEffective2024,Kim_2024_extraordinary,klattTransparencyAndersonLocalization2026}.
Here, we address this intriguing theoretical problem with a minimal 1D continuum model, namely the Kronig--Penney-like model.
Our conclusions not only offer an in-depth expansion for previous numerical simulation observations, but also are relevant for applications to the future design of transparent disordered materials.

Hyperuniform systems are characterized by an anomalous suppression of long-range density fluctuations compared to those of conventional disordered systems~\cite{To18a,To03a}. Focusing on $d$-dimensional point patterns, whether a given system is hyperuniform can be determined by computing the number variance $\sigma^2_N(R) = \langle N^2(R) \rangle - \langle N(R) \rangle^2$ of points contained in an observation window of linear size $R$. If the ratio $\sigma^2_N(R)/R^d$ vanishes as $R \to \infty$, the system is hyperuniform. Equivalently, in a hyperuniform system, the structure factor $S(k)$ approaches $0$ as $k \to 0$. 

Stealthy hyperuniform (SHU) systems represent a subclass of hyperuniform configurations in which the structure factor $S(k)$ vanishes in a range of wave numbers $0 \le k < K$ around the origin~\cite{uche_constraints_2004,batten_classical_2008,Zh15a,Zh15b}. While crystals are an example of SHU systems, disordered configurations belonging to this class are endowed with intermediate liquid-crystal properties, such as the absence of arbitrarily large holes~\cite{Gh18} and the absence of single scattering from intermediate to infinite wavelengths~\cite{To21a}, yet remaining isotropic and without Bragg peaks.

According to Fermi's golden rule \cite{sakuraiModernQuantumMechanics2020}, the lowest-order single-scattering contribution is dominated by the structure factor $S(q)$ of the momentum transfer $q$, e.g., $q=2k$ in a 1D scattering between $k$ and $-k$. Therefore, the SHU correlations can strongly modify the localization properties of one-dimensional models~\cite{jeon2026,koga2026}.
For instance, it has recently been found~\cite{vanoni2025effective} that the 1D Anderson model with stealthy Gaussian disorder displays effectively delocalized states. More specifically, for fixed energy $E$ and weak disorder strength $W$, the stealthy cutoff $K$ can be chosen so that the leading nonvanishing contribution to the inverse localization length appears only at arbitrarily high order in perturbation theory. Equivalently, the localization length scales as $\xi \sim W^{-2m}$ with an arbitrarily large integer $m$, differently from the $\xi \sim W^{-2}$ scaling with uncorrelated disorder. Thus, for any finite system size $L$, one can obtain regimes in which $\xi>L$, producing a finite interval of effectively extended states.

When generalized from the discrete random point patterns to the continuous two-phase random media, stealthy hyperuniformity also affects the electromagnetic transmission through disordered stealthy hyperuniform layered media composed of alternating high- and low-index slabs~\cite{kim_effective_2023,kimTheoreticalPredictionEffective2024,Kim_2024_extraordinary}. Within the nonlocal strong-contrast expansion~\cite{To21a}, the effective dynamic dielectric constant has been shown to have a vanishing imaginary part through third order, implying perturbative transparency in the stealthy regime $0\le k<K/2$. In addition, transfer matrix numerical calculations of the Lyapunov exponent~\cite{klattTransparencyAndersonLocalization2026} have not detected signs of localization at small wave numbers $0\le k<K/2$ for systems up to $L=10^4$. It remains unclear, however, whether this transparency persists in the thermodynamic limit and to all orders in perturbation theory, or whether the apparent transparency represents an effective delocalization regime analogous to that found in the Anderson model.

In this paper, we study the 1D propagation of a quantum particle in the presence of disordered $\delta$-function scatterers subject to the stealthy hyperuniform constraint.
This can be viewed as a generalized form of the Kronig--Penney (KP) model---which gives one of the simplest derivations of Bloch bands and band gaps---by simply replacing the periodic array with a correlated disordered point pattern.
Unlike the standard Anderson model, in which a particle hops on a lattice with random on-site energies, KP and KP-like models describe continuum propagation subject to scattering from localized potentials. 

We analytically compute the Lyapunov exponent $\lambda(k)$ (inverse localization length $\xi^{-1}$) in the thermodynamic limit up to the fourth order in amplitude of the potential $W$ under free boundary conditions, and its counterpart (inverse lifetime $\tau^{-1}$) under periodic boundary conditions. Within the stealthy range $0\le k<K/2$, we find that the disorder-dependent corrections to the Lyapunov exponent $\lambda(k)$ vanish through third order in $W$, whereas a small but nonzero contribution appears at fourth order. This conclusion agrees with what is observed in stealthy layered media~\cite{kim_effective_2023,kimTheoreticalPredictionEffective2024,Kim_2024_extraordinary,klattTransparencyAndersonLocalization2026}, where transparency also persists at least through the third order in terms of the dielectric contrast of the two-phase medium.
We further show that, in contrast with the Anderson model with Gaussian disorder, large transmittance $T(k)=\mathrm{e}^{-\lambda(k)L}$, or equivalently, small Lyapunov exponent $\lambda(k)$ in the KP-like model requires not only the suppression of scattering over a finite stealthy interval $0\le k<K/2$, but also the hyperuniform suppression of large-scale density fluctuations. This distinction originates from the geometric nature of the disorder, which enters through the scatterer positions rather than through independent potential amplitudes. 

In this KP-like model, we show that the stealthy correlations yield a Lyapunov exponent $\lambda(k)\sim O(W^4)$ rather than the standard $O(W^2)$ in the scattering strength $W$ for $k$ below a certain cutoff, and obtain an explicit and simple form of the $O(W^4)$ dependence.
Based on this expression, we demonstrate how the hyperuniformity condition further reduces this fourth-order residual $O(W^4)$ and enhances the delocalization effect, filling the gap between prior perturbation theory and simulations~\cite{vanoni2025effective,kim_effective_2023,kimTheoreticalPredictionEffective2024,Kim_2024_extraordinary,klattTransparencyAndersonLocalization2026,Tessieri2002Delocalization}.

\textbf{KP-like model and stealthy hyperuniformity.---}
We consider the propagation of a quantum particle in one dimension in the presence of $N$ identical point scatterers located at positions $0\le x_1\le x_2\le\ldots\le x_N<L$. The potential $V(x)$ in the interval $[0,L)$ is given by
\begin{equation}
\label{eq:potential}
    V(x)=Wn_{_N}(x)= W\sum_{n=1}^N\delta(x-x_n),
\end{equation}
where $W$ denotes the scattering strength and $n_{_N}(x)\equiv\sum\limits_{n=1}^N\delta(x-x_n)$ denotes the local density of scatterers. 
We derive the expansion of the Lyapunov exponent $\lambda=1/\xi$ (inverse localization length) under free boundary conditions and the inverse lifetime $1/\tau$ under periodic boundary conditions; the former assumes free space outside the range $0\le x<L$, so that the full potential is simply $U^{\text{fbc}}(x)=V(x)$, while the latter assumes periodic images of the potential $V(x)$ within the range $0\le x<L$, i.e., $U^{\text{pbc}}(x)=\sum \limits_{m=-\infty}^{\infty} V(x-mL)$. We will discuss the free-boundary-condition case in the main text, while reporting the periodic-boundary-condition results and their agreement with free-boundary-condition ones in the Supplemental Material.
For simplicity, we work at unit density $\rho \equiv N/L = 1$ to simplify the notation in the finite-$N$ calculations. 

\begin{figure}
    \centering
    \includegraphics[width=\linewidth]{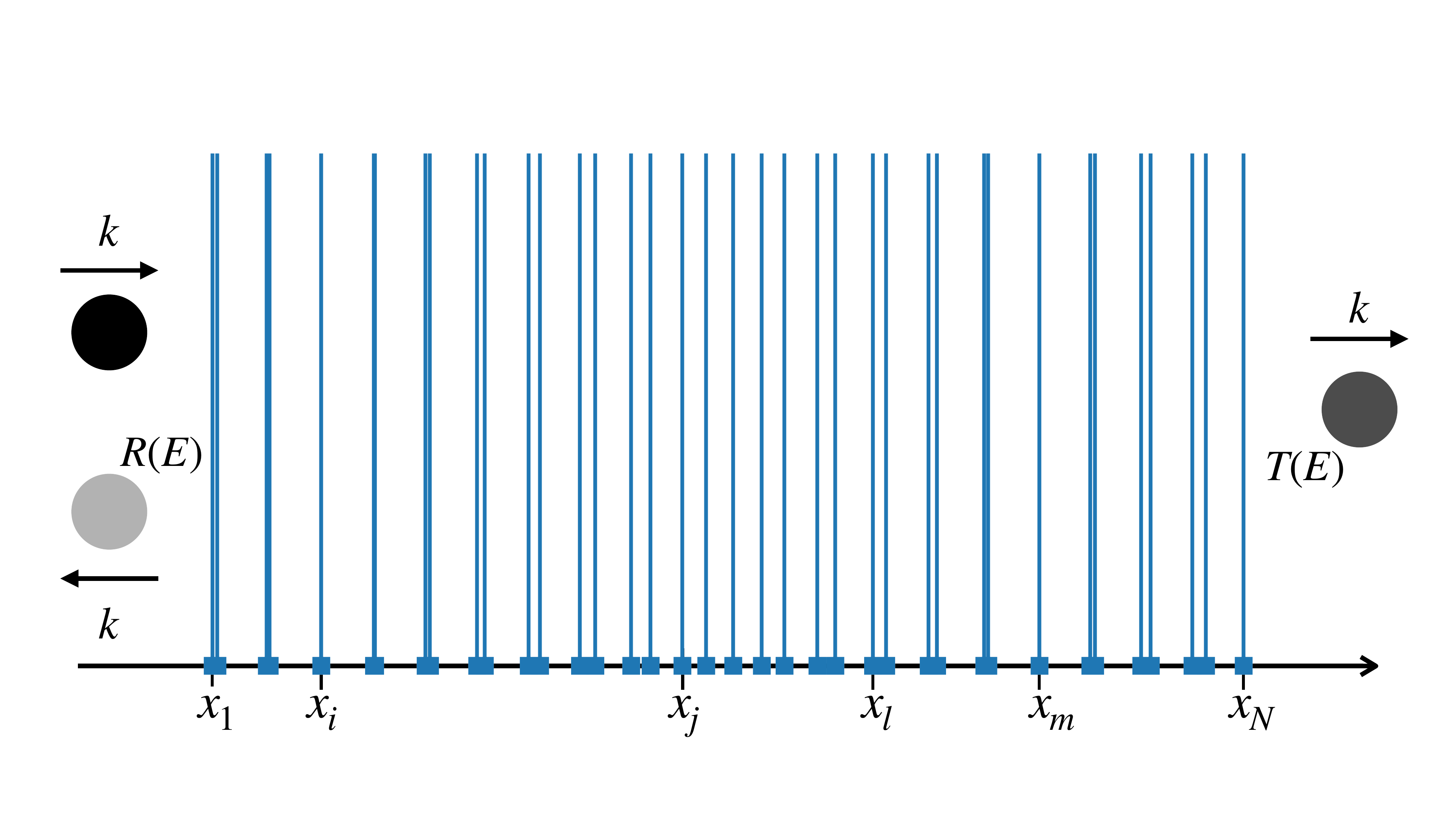}
    \caption{Schematic of the quantum particle transmission through the disordered potentials $V(x)$ in Eq.~\eqref{eq:potential}. A particle coming from $x = -\infty$ with momentum $k$ and energy $E=k^2$ (black) scatters against $N$ $\delta$-functions, and is transmitted with probability $T(E)$, while it is reflected with probability $R(E)$.}
    \label{fig:schematic}
\end{figure}

We consider the propagation of a quantum particle of energy $E=k^2$ moving in the positive-$x$ direction, as depicted schematically in Fig.~\ref{fig:schematic}.
Our goal is to compute the probability to find the particle at position $x>L$, that is, the transmittance $T(E)$ of the potential $V(x)$. The transmittance is related to the reflectance $R(E)$, the probability of the particle getting reflected by the potential, via the sum rule $T(E) + R(E) = 1$.
In generic KP-like models, it can be shown~\cite{gasparianResistanceOnedimensionalChains1988} that the transmittance $T_N(E)$ is given by
\begin{equation}
    T_N(E)=|D_N(E)|^{-2} \equiv |\det(\mathbb{I}+W\mathbb{G})|^{-2},
\end{equation}
where $\mathbb{I}$ is the $N\times N$ identity matrix, and $\mathbb{G} = [G_{ij}]$, with $G_{ij} = \mathrm{i}\,\mathrm{e}^{\mathrm{i}k|x_{ij}|}/2k $ being the free-particle Green's function, and $x_{ij}\equiv x_i-x_j$.
The Lyapunov exponent $\lambda_{_N}(k)$ is defined as
\begin{equation}
\label{eq:lambda_T}
    \lambda_{_N}(k)\equiv -\frac{1}{N}\ln[T_{N}(E)]=\frac{2}{N}\Re[\ln{D_N(E)}],
\end{equation}
whose thermodynamic limit $\lambda(k)\equiv \lim\limits_{N \to \infty}\expval{\lambda_{_N}(k)} $ corresponds to the inverse of the localization length $\xi$.

As motivated in the introduction, we are interested in studying the effect of stealthy hyperuniformity on the transport properties of this model. Let us define the scattering intensity
\begin{equation}
\label{eq:S_N}
    \mathscr{S}_N(k)\equiv\frac{1}{N}\sum\limits_{i,j=1}^N\mathrm{e}^{\mathrm{i}k \, x_{ij}}=\frac{1}{N}|\tilde{n}_{_N}(k)|^2,
\end{equation}
with $\tilde{n}_{_N}(k)\equiv\sum\limits_{n=1}^N\mathrm{e}^{-\mathrm{i}kx_n}$ being the Fourier transform of the local density of scatterers $n_{_N}(x)$. The thermodynamic limit of the scattering intensity $\mathscr{S}_N(k)$ is related to the structure factor $S(k)$ via 
\begin{equation}
\label{eq:S_TDL}
    \mathscr{S}(k)\equiv \lim_{N \to \infty}\expval{\mathscr{S}_N(k)}=S(k)+2\pi\delta(k).
\end{equation}
We will consider the positions $x_i$ of the scatterers to satisfy the SHU condition, namely $S(k) = 0$ for $0\leq k < K$. 
Practically, one-dimensional SHU point configurations can be obtained using the Collective Coordinate Optimization procedure described in Refs.~\cite{uche_collective_2006,To15,zhang_can_2017}.

\begin{figure}
    \centering
    \subfloat[$n=1$]{%
        \includegraphics[width=0.4\linewidth,page=1]{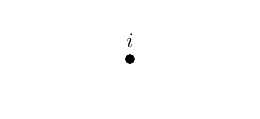}%
        \label{fig:G1}%
    }
    \subfloat[$n=2$]{%
        \includegraphics[width=0.4\linewidth,page=2]{fig2.pdf}%
        \label{fig:G2}%
    }\\
    \subfloat[$n=3$]{%
        \includegraphics[width=0.4\linewidth,page=3]{fig2.pdf}%
        \label{fig:G3}%
    }
    \subfloat[$n=4$]{%
        \includegraphics[width=0.4\linewidth,page=4]{fig2.pdf}%
        \label{fig:G4}%
    }
    \caption{Feynman diagrams of the first 4 orders under free boundary conditions. Each vertex (e.g., $i$) contributes a factor of $\frac{\mathrm{i}}{2k}$, and each edge between two vertices, e.g., $i,j$, contributes a factor of $\mathrm{e}^{\mathrm{i}|x_{ij}|}\equiv\mathrm{e}^{\mathrm{i}|x_{ij}|}$, where we denote $x_{ij}\equiv x_i-x_j$. When evaluated by the summation over each vertex, e.g., $i$, from $i=1$ to $i=N$, the $n$-th-order diagram gives $\Tr[\mathbb{G}^n]$.}
    \label{fig:feynman_fbc}
\end{figure}

\textbf{Perturbative calculation.---}
We now compute the perturbative corrections to the Lyapunov exponent $\lambda_{_N}(k)$ for small $W$ under free boundary conditions. We start by expanding Eq.~\eqref{eq:lambda_T}
\begin{equation}
    \begin{aligned}
        \lambda_{_N}(k)&=\frac{2}{N}\Re[\ln{D_N(E)}]\\
        &=\frac{2}{N}\sum_{n=1}^\infty\frac{(-1)^{n-1}}{n}W^n\Re[\Tr[\mathbb{G}^n]],
    \end{aligned}
\end{equation}
where we used the trace equivalence $\ln \det (\cdot) = \Tr \ln (\cdot)$ and expanded the logarithm for small $W$.
Therefore, the computation of the $O(W^n)$ contribution to the Lyapunov exponent boils down to computing the real part of $\Tr[\mathbb{G}^n]$. Physically, this quantity corresponds to the $n$-point scattering processes, as depicted in Fig.~\ref{fig:feynman_fbc}. 

We now derive the contributions to $\lambda(k)$ through fourth order in $W$, identifying the scattering processes and structural correlations that control localization at each order. In the following, we will denote with a subscript $N$ functions of the finite-sized system, while the absence of the subscript denotes the disorder average in the thermodynamic limit, i.e., $F(k) \equiv \lim\limits_{N \to \infty} \langle F_N(k)\rangle$.

\paragraph{First order.} The first-order contribution is simply given by the trace of the propagator, $\Tr\mathbb{G}=\mathrm{i}\frac{N}{2k}$ [Fig.~\ref{fig:G1}], and is purely imaginary. Physically, it only adds to a phase shift and does not affect the transmittance $T_N$. 

\paragraph{Second order.} The computation of the second-order contribution [Fig.~\ref{fig:G2}] is straightforward and gives
\begin{equation}
\label{eq:TrG2}
    \Tr[\mathbb{G}^2]=-\frac{1}{4k^2}\sum_{i,j=1}^N\mathrm{e}^{2\mathrm{i}k|x_{ij}|}\equiv-\frac{N}{4k^2}Z_N(2k),
\end{equation}
with $Z_N(q)$ defined as Eq.~\eqref{eq:ZN_def} such that $\Re[Z_N(q)] = \mathscr{S}_N(q)$. 
In the thermodynamic limit, $Z(q)$ has real part $\Re[Z(q)]=\mathscr{S}(q)=S(q)$ assuming $q\neq 0$, and thus the second-order contribution to the Lyapunov exponent is proportional to the structure factor. 
The SHU condition then implies the transmittance is unity at second order if $k<K/2$, as expected by the suppression of direct backscattering.

\paragraph{Third order.} The third order contribution [Fig.~\ref{fig:G3}] is less trivial, and it can be shown to be
\begin{equation}\label{eq:order3}
    \begin{aligned}
        \Tr[\mathbb{G}^3]&=-\frac{\mathrm{i}}{8k^3}\sum_{i,j,l=1}^N\mathrm{e}^{\mathrm{i}k(|x_{ij}|+|x_{jl}|+|x_{li}|)}\\
        &=-\frac{\mathrm{i}N}{8k^3}\left[1+3J_N(2k)\right],
    \end{aligned}
\end{equation}
where the constant term arises from coincident indices $i=j=l$, while $J_N$, defined in Eq.~\eqref{eq:JN_def}, collects all remaining triangular paths.
In one dimension, a closed triangular path $i\to j\to l\to i$ necessarily contains two backscattering events and one forward-scattering event.
As a result, the third-order contribution depends only on two-point structural information, rather than on an independent three-point correlation function.
Noting that the factor $|i-j|$ in $J_N$ equals the number of points $N_p(D)$ within the separation $D=|x_{ij}|$. For a statistically homogeneous disordered system of unit density, its ensemble average must be given by $\expval{N_p(D)}=D$. Hyperuniformity further suppresses the fluctuations around this mean~\cite{Vanoni2025Quantifying}, improving the convergence for individual realizations.
Therefore, in the thermodynamic limit, Eq.~\eqref{eq:JN_def} becomes $J(q)=-\mathrm{i} Z'(q)$ and thus the contribution to the Lyapunov exponent $\lambda(k)$ is proportional to $\Im[J(2k)]=-S'(2k)$, which vanishes for $k<K/2$ for SHU positions of the scatterers. 

\paragraph{Fourth order.} The fourth order [Fig.~\ref{fig:G4}] is more complicated, as can be expressed as
\begin{equation}
    \begin{aligned}
        \Tr[\mathbb{G}^4]&=\frac{1}{16k^4}\sum_{i,j,l,m=1}^N\mathrm{e}^{\mathrm{i}k(|x_{ij}|+|x_{jl}|+|x_{lm}|+|x_{mi}|)}\\
        &=\frac{N}{16k^4}[4H_N(2k)+2Q_N(2k)+Z_N(4k)],
    \end{aligned}
\end{equation}
where the functions $H_N$ and $Q_N$ are defined in Eqs.~\eqref{eq:HN_def} and~\eqref{eq:QN_def} of the End Matter, where we detail how they arise from considering the possible combinations of four-scatterer paths in one dimension.

Combining the contributions through fourth order gives, in the thermodynamic limit, for any disordered 1D pattern, the Lyapunov exponent expansion
\begin{equation}\label{eq:exact_expan}
    \begin{aligned}
        \lambda(k)={}&\frac{W^2}{4k^2}S(2k)-\frac{W^3}{4k^3}S'(2k)\\
        &-\frac{W^4}{32k^4}\Re[4H(2k)+2Q(2k)+S(4k)]+O(W^5).
    \end{aligned}
\end{equation}
Next, our goal is to further decompose $H(q)$ and $Q(q)$ in terms of the structure factor $S(q)$ and its derivatives and integrals.
In particular, we are interested in estimating the fourth-order coefficient of disordered SHU systems in the effectively transparent regime $k<K/2=\pi\chi$, where the second- and third-order coefficients vanish, and thus the fourth-order contribution becomes the leading order.

As for $Q(q)$, motivated by the the Kirkwood superposition approximation (KSA)~\cite{kirkwoodStatisticalMechanicsFluid1935}, we can ignore the irreducible four-point correlation and keep only the trivial part given by products of lower-order correlation functions~\footnote{Mathematically, $\sum \limits_{l,m}^{}\mathrm{e}^{\mathrm{i}q|x_{lm}|}\approx|i-j|\times Z(q)$ in Eq.~\eqref{eq:HN_def}.}. When doing this, we get the simpler relation $Q(q)\approx J(q)Z(q)$.

As for $H(q)$, we can introduce the auxiliary quantity $\Omega_N(q)\equiv H_N(q)+Z''(q)$ (see Eq.~\eqref{eq:OmegaN_def}), thus $\Re[H(q)]\equiv \Re[\Omega(q)]-S''(q)$.
On the integer lattice $|i-j|=|x_{ij}|$, implying $\Omega_N(q)\equiv0$. 
For disordered systems, instead, the factor $(i-j)^2-(x_{ij})^2=[N_p(D)]^2-D^2$ in Eq.~\eqref{eq:OmegaN_def} approaches the number variance $\sigma^2_{D}$ within the same separation $D=|x_{ij}|$ in the ensemble-average sense, unless the presence of points at the boundary strongly affects the statistics in between.
Specifically, for disordered SHU systems, this variance remains bounded at the long-distance limit, approaching the constant~\cite{torquatoStructuralCharacterizationManyparticle2021}
\begin{equation}\label{eq:c}
    c=\lim\limits_{D\to\infty} \sigma^2_{D}=\frac{2}{\pi}\int_{2\pi\chi}^\infty\frac{\mathrm{d}q}{q^2}S(q).
\end{equation}
In the exclusion region $q<K=2\pi\chi$, we get $\Omega(q)=c[Z(q)-1]$ and thus $\Re[\Omega(q)]=-c$.
Meanwhile, we have $\Re[Q(q)]\approx \Re[J(q)Z(q)]=\mathrm{i}\,\Re[\Im[Z'(q)]\Im[Z(q)]]=0$, as $\Im[J(q)]=-S'(q)=0$ and $\Re[Z(q)]=S(q)=0$.

In summary, up to the order of $W^4$ and in the thermodynamic limit, for disordered SHU systems within the effectively transparent regime $k<K/2=\pi\chi$, we have
\begin{equation}
    \begin{aligned}
        \lambda(k)&=-\frac{W^4}{32k^4}\Re[4\Omega(2k)+2Q(2k)+S(4k)]+O(W^5)\\
        &\approx\frac{W^4}{32k^4}[4c-S(4k)]+O(W^5),
    \end{aligned}
    \label{eq:4th_order}
\end{equation}
where the first line is exact, while in the second line we plug in the approximation $\Re[Q(2k)]\approx\Re[J(2k)Z(2k)]=0$ as just explained.

\textbf{Analysis of the perturbation theory result.---}
We have, for the first time to our knowledge, derived the simplified exact perturbative expression~\eqref{eq:exact_expan} of the Lyapunov exponent $\lambda(k)$ for a continuum model of disorder (namely, the Kronig--Penney-like model) up to fourth order. 
This perturbative expansion is further simplified into Eq.~\eqref{eq:4th_order} for disordered SHU systems in the effectively transparent interval $k<K/2=\pi\chi$, where the fourth order rather than the second order becomes the leading order. 
Additionally, in the Supplemental Material, we demonstrate the perturbative calculation of the inverse lifetime $1/\tau(k)$ under periodic boundary conditions, in parallel to and perfectly matching that of the Lyapunov exponent $\lambda(k)$ under free boundary conditions.
Our calculations indicate that both stealthiness ($S(q)=0$ for a certain range of $q$) and hyperuniformity ($S(q\to0)=0$) are necessary conditions for enhancing the effective transparency, i.e., minimizing $\lambda(k)$ for a certain window of $k$. 
Without stealthiness, the second- and third-order terms in the expansion~\eqref{eq:exact_expan} will not vanish.
Without hyperuniformity, $\Omega(2k)$ and thus the fourth-order contribution in $\lambda(k)$ will become much larger due to the increased number variance $\sigma^2_D$.

Our results bridge the gap with previous work in the literature.
Tessieri \cite{Tessieri2002Delocalization} reported the vanishing second- and third-order corrections in the Lyapunov exponent $\lambda(k)$ for a 1D quantum disordered system with stealthy but nonhyperuniform two-point correlation, but failed to quantitatively estimate what he referred to as the ``non-Gaussian'' part of the fourth-order term.
Kim and Torquato~\cite{kim_effective_2023,kimTheoreticalPredictionEffective2024,Kim_2024_extraordinary} demonstrated the effective transparency of disordered SHU two-phase media by both theoretically deriving the vanishing second- and third-order corrections within the strong contrast expansion and numerically simulating the electromagnetic wave propagation, but the existence of fourth- and higher-order corrections remained an open question. 
The 1D simulation of the EM wave propagation in disordered SHU media on a larger scale conducted by Klatt \textit{et al.}~\cite{klattTransparencyAndersonLocalization2026} could not determine whether $\lambda(k)$ is exactly zero or too small to be detected in the SHU effective transparency region.
Though our Eq.~\eqref{eq:exact_expan} and Eq.~\eqref{eq:4th_order} are derived from a different problem setting from \cite{kim_effective_2023,kimTheoreticalPredictionEffective2024,Kim_2024_extraordinary,klattTransparencyAndersonLocalization2026}, we can still estimate the order of magnitude semi-quantitatively with them. 
Plugging in $W\sim\beta\phi$, where $\beta\sim1$ is the strong-contrast expansion parameter and $\phi=0.01$ is the volume fraction~\cite{klattTransparencyAndersonLocalization2026}, with the fourth-order coefficient $\sim1$ as given by Eq.~\eqref{eq:4th_order}, the nonzero residual of $\lambda(k)$ should be of the order of $10^{-8}$ ($\rho=1$), smaller than the numerical error induced by the ``ripple effect''~\cite{klattTransparencyAndersonLocalization2026}, and thus not detectable in the simulations.

\textbf{Conclusions.---}
In summary, we have isolated the structural mechanism underlying the apparent transparency of one-dimensional stealthy hyperuniform media. Using a minimal continuum model of identical point scatterers, we have shown that stealthiness exactly eliminates the contributions to the Lyapunov exponent $\lambda(k)$ through third order in the scattering strength in the interval $0<k<K/2$. The leading attenuation is therefore shifted from the conventional $O(W^2)$ scaling of one-dimensional disordered systems to $O(W^4)$. Localization is not generically eliminated in the thermodynamic limit; instead, it is generated only by higher-order multiple-scattering processes. Consequently, the localization length $\xi\sim1/\lambda\sim W^{-4}$ becomes parametrically large at weak scattering, producing a broad regime in which finite samples behave as effectively transparent even though their asymptotic eigenstates remain localized.

Stealthiness and hyperuniformity play distinct but complementary physical roles: stealthiness eliminates low-order backscattering, whereas hyperuniformity bounds large-scale number fluctuations and suppresses the residual fourth-order attenuation. For parameters comparable to recent SHU layered-media calculations, our estimate $\lambda \sim 10^{-8}$ places this attenuation below the reported transfer-matrix resolution and is consistent with their observed effective transparency. These results connect quantum localization, classical-wave propagation, and structural statistics, establishing SHU positional order as a broadly applicable design principle for disordered media with parametrically enhanced transmission.

\textbf{Data availability.---}
The data that support the findings of this article are not publicly available because they are proprietary. The data are available from the authors upon reasonable request.

\textbf{Acknowledgments.---}
We are grateful to Boris L. Altshuler for his useful suggestions and collaboration on related topics.
This work was supported by the Army Research Office under Cooperative Agreement No. W911NF-22-2-0103.

%


\begin{center}
{\large \bf End Matter}\\
\end{center}
\twocolumngrid

We report here the expressions and meanings of the functions $Z_N(k)$, $J_N(k)$, $H_N(k)$, $Q_N(k)$, and $\Omega_N(k)$ that arise in the perturbative calculation in the main text.

In the second-order calculation, we introduce
\begin{equation}
\label{eq:ZN_def}
    \begin{aligned}
        Z_N(q)&\equiv \frac{1}{N} \sum\limits_{i,j=1}^N\mathrm{e}^{\mathrm{i}q|x_{ij}|}\equiv \mathscr{S}_N(q)+\frac{\mathrm{i}}{N} \sum\limits_{i,j=1}^N\sin(q|x_{ij}|)\\
        &=\mathscr{S}_N(q)-2\mathrm{i}\int_{-\infty}^\infty\frac{\mathrm{d}p}{2\pi}\frac{\mathscr{S}_N(q)-\mathscr{S}_N(p)}{q-p},
    \end{aligned}
\end{equation}
so that $\Re[Z_N(q)] = \mathscr{S}_N(q)$.

In the third-order calculation, noting that $|x_{ij}|+|x_{jl}|+|x_{li}|\equiv2|x_{\max(i,j,l)}-x_{\min(i,j,l)}|$, independent of the point scatterer located between the other two, we introduce the simplified double summation
\begin{equation}
\label{eq:JN_def}
    J_N(q)\equiv \frac{1}{N} \sum\limits_{i,j=1}^N |i-j|\mathrm{e}^{\mathrm{i}q|x_{ij}|}.
\end{equation}
Geometrically, this simplification comes from the aforementioned 1D triangular equality.

For the fourth order, supposing $\{i,j,l,m\}=\{i_1,i_2,i_3,i_4\}$ with $i_1\le i_2\le i_3\le i_4$, the summation $|x_{ij}|+|x_{jl}|+|x_{lm}|+|x_{mi}|$ can equal either $2|x_{i_1i_4}|$ or $2|x_{i_1i_4}|+2|x_{i_2i_3}|$, depending on the permutation of indices.
Roughly, the former gives the $H_N(2k)$ term, while the latter gives both $Q_N(2k)$ ($i_1\ne i_2$ or $i_3\ne i_4$) and $Z_N(4k)$ ($i_1=i_2$ and $i_3=i_4$), where we define
\begin{equation}\label{eq:HN_def}
    H_N(q)\equiv\frac{1}{N}\sum_{i\ne j=1}^N(i-j)^2\mathrm{e}^{\mathrm{i}q|x_{ij}|},
\end{equation}
\begin{equation}\label{eq:QN_def}
    \begin{aligned}
        Q_N(q) \equiv{} & \frac{1}{N}\sum_{i, j=1}^N\mathrm{e}^{\mathrm{i}q|x_{ij}|}\times\\
        &{}\frac{1}{2}\left(\sum_{l,m=\min(i,j)}^{\max(i,j)-1}+\sum_{l,m=\min(i,j)+1}^{\max(i,j)}\right)\mathrm{e}^{\mathrm{i}q|x_{lm}|},
    \end{aligned}
\end{equation}
respectively.
Physically, $H_N(2k)$ corresponds to particle trajectories containing two backscattering and two forward-scattering events, while $Q_N(2k)$ and $Z_N(4k)$ correspond to those with four backscattering events.
In particular, $Z_N(4k)$ describes repeated bouncing between the same pair of scatterers, whereas $Q_N(2k)$ includes processes involving at least three distinct scatterers.
Thus, $Q_N$ contains genuine four-point correlation information, while $H_N$ and $Z_N$ depend only on two-point quantities. 
Finally, noting that $|i-j|\sim|x_{ij}|$, we introduce
\widetext
\begin{equation}
\label{eq:OmegaN_def}
\Omega_N(q) \equiv \frac{1}{N}\sum_{i\ne j=1}^N[(i-j)^2-(x_{ij})^2]\mathrm{e}^{\mathrm{i}q|x_{ij}|} \equiv H_N(q)+Z_N''(q).
\end{equation}


\section*{--- Supplemental Material ---}

We now discuss the perturbative calculations of the same Kronig–Penney-like model under periodic boundary conditions. In this setting, it is more natural to work with the lifetime $\tau$ rather than the localization length $\xi$. The periodic potential is given by
\begin{equation}
    U^{\text{pbc}}(x)=\sum \limits_{m=-\infty}^{\infty} V(x-mN)=\frac{1}{N}\sum_{q}\tilde{V}_q\mathrm{e}^{\mathrm{i}qx},
\end{equation}
where $q=0,\pm\frac{2\pi}{N},\pm\frac{4\pi}{N},\ldots$ and $\tilde{V}_q\equiv W\tilde{n}_{_N}(q)$. The Schr\"odinger equation in the Fourier space is given by
\begin{equation}
    \mathrm{i}\partial_t\psi_k=k^2\psi_k+\frac{1}{N}\sum_{q}\tilde{V}_{k-q}\psi_{q},
\end{equation}
where we set $\hbar=1$, and the effective eigenfrequency $\omega_{_N}(k)$ can be formally expanded as
\begin{equation}
    \omega_{_N}(k)=k^2+W\left[1+\sum_{m=2}^\infty W^{m-1}A_N^{(m)}(k)\right]
\end{equation}
according to quantum perturbation theory. The thermodynamic limit of the imaginary part of $\omega_{_N}(k)$ corresponds to the inverse of the lifetime $\tau$.

\begin{figure}
    \centering
    \subfloat[$A_N^{(2)}(k)=\mathcal{A}_N^{(2)}(k)$]{%
        \includegraphics[width=0.25\linewidth,page=1]{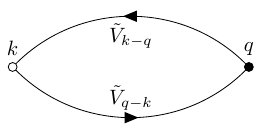}%
        \label{fig:A2}%
    }
    \subfloat[$A_N^{(3)}(k)$ or $\mathcal{A}_N^{(3)}(k)$]{%
        \includegraphics[width=0.25\linewidth,page=2]{fig3.pdf}%
        \label{fig:A3}%
    }
    \subfloat[$B_N^{(3)}(k)=\mathcal{B}_N^{(3)}(k)$]{%
        \includegraphics[width=0.25\linewidth,page=3]{fig3.pdf}%
        \label{fig:B3}%
    }
    \subfloat[$A_N^{(4)}(k)$ or $\mathcal{A}_N^{(4)}(k)$]{%
        \includegraphics[width=0.25\linewidth,page=4]{fig3.pdf}%
        \label{fig:A4}%
    }\\
    \subfloat[$B_N^{(4a)}(k)$ or $\mathcal{B}_N^{(4a)}(k)$]{%
        \includegraphics[width=0.25\linewidth,page=5]{fig3.pdf}%
        \label{fig:B4a}%
    }
    \subfloat[$B_N^{(4b)}(k)$ or $\mathcal{B}_N^{(4b)}(k)$]{%
        \includegraphics[width=0.25\linewidth,page=6]{fig3.pdf}%
        \label{fig:B4b}%
    }
    \subfloat[$B_N^{(4c)}(k)$ or $\mathcal{B}_N^{(4c)}(k)$]{%
        \includegraphics[width=0.25\linewidth,page=7]{fig3.pdf}%
        \label{fig:B4c}%
    }
    \subfloat[$C_N^{(4)}(k)=\mathcal{C}_N^{(4)}(k)$]{%
        \includegraphics[width=0.25\linewidth,page=8]{fig3.pdf}%
        \label{fig:C4}%
    }
    \caption{Feynman diagrams of the first 4 orders under periodic boundary conditions. $A_N$ and $\mathcal{A}_N$ denote 1-loop diagrams, $B_N$ and $\mathcal{B}_N$ denote 2-loop diagrams, and so on. For a fixed $k$, $A_N,B_N,\ldots$ are given by evaluating the diagrams under the self-avoiding summation $\sum_{q_1\ne q_2\ne\ldots\ne q_{m-1}\ne k}$ over the inner momenta $q_1,q_2,\ldots,q_{m-1}$ (some of which are renamed as $q,p$ in the diagrams), while $\mathcal{A}_N,\mathcal{B}_N,\ldots$ by evaluating the same diagrams under the non-self-avoiding summation $\sum_{q_1,q_2,\ldots,q_{m-1}\ne k}$. $A_N,B_N,\ldots$ and $\mathcal{A}_N,\mathcal{B}_N,\ldots$ of the same order are related to each other by a linear transformation.}
    \label{fig:feynman_pbc}
\end{figure}

The effective eigenfrequency $\omega_{_N}(k)$ can be formally expanded as

\begin{align}
    \omega_{_N}(k)&=k^2+\frac{1}{N}\tilde{V}_{k-k}+\frac{1}{N^2}\sum_{q\ne k}\frac{\tilde{V}_{k-q}\tilde{V}_{q-k}}{k^2-q^2+\mathrm{i}0^+}
    +\frac{1}{N^3}\sum_{q_1\ne q_2\ne k}\frac{\tilde{V}_{k-q_1}\tilde{V}_{q_1-q_2}\tilde{V}_{q_2-k}}{(k^2-q_1^2+\mathrm{i}0^+)(k^2-q_2^2+\mathrm{i}0^+)}+\ldots \\
    &\equiv k^2+W \left[1+\sum_{m=2}^\infty W^{m-1}A_N^{(m)}(k)\right],
\end{align}

where we define the coefficients as
\begin{equation}
A_N^{(m)}(k)\equiv\frac{1}{ N^m}\sum_{q_1\ne q_2\ne\ldots\ne q_{m-1}\ne k}\tilde{n}_{_N}(k-q_1) \frac{\left[\prod_{i=1}^{m-2}\tilde{n}_{_N}(q_{i}-q_{i+1})\right]\tilde{n}_{_N}(q_{m-1}-k)}{\prod_{i=1}^{m-1}(k^2-q_i^2+\mathrm{i}0^+)}.
\end{equation}
To simplify the expression, we define
\begin{equation}
    \begin{aligned}
        \mathcal{A}_N^{(m)}(k)&\equiv\frac{1}{ N^m}\sum_{q_1,q_2,\ldots,q_{m-1}\ne k}\tilde{n}_{_N}(k-q_1)
        \frac{\left[\prod_{i=1}^{m-2}\tilde{n}_{_N}(q_{i}-q_{i+1})\right]\tilde{n}_{_N}(q_{m-1}-k)}{\prod_{i=1}^{m-1}(k^2-q_i^2+\mathrm{i}0^+)}\\
        &\to\frac{1}{N}\left(\frac{-\mathrm{i}}{4k}\right)^{m-1}\sum_{i_1,i_2,\ldots,i_{m}=1}^N\prod_{p=1}^{m-1}\left[2\mathrm{e}^{2\mathrm{i}k(x_{i_p}-x_{i_{p+1}})_+}-1\right],
    \end{aligned}
\end{equation}
where we use the substitution 
\begin{equation}
    \frac{1}{N}\sum_{q_i\ne k}\to\left(\int_{-\infty}^{k_-}+\int_{k_+}^\infty\right)\frac{\mathrm{d}q_i}{2\pi}
\end{equation}
as $N\to\infty$, and we define $x_+\equiv\frac{|x|+x}{2}$.
Notice that
\begin{equation}
    \sum_{q_1\ne q_2\ne\ldots\ne q_{m-1}\ne k}\equiv\sum_{q_1,q_2,\ldots,q_{m-1}\ne k}-\sum_{\substack{q_1,q_2,\ldots,q_{m-1}\ne k,\\\exists q_i=q_j}},
\end{equation}
so we have
\begin{equation}
    A_N^{(m)}(k)\equiv\mathcal{A}_N^{(m)}(k)-\text{high-loop diagram terms}.
\end{equation}
The high-loop diagram terms reflect the forward scattering effects, and thus contain factors of  $N$ or powers of $N$ and diverge in the thermodynamic limit $N\to\infty$. After removing these divergent terms, what remains should be a physically meaningful one-loop coefficient $A_N^{(m)}(k)$ that has a finite thermodynamic limit $A^{(m)}(k)=\lim\limits_{N \to \infty}\expval{A_N^{(m)}(k)}$. Specifically, the Feynman diagrams of the first 4 orders are presented in Fig.~\ref{fig:feynman_pbc}.

The imaginary part of $\omega_{_N}(k)$ is related to the lifetime $\tau_{_N}(k)$ by
\begin{equation}
    \frac{1}{\tau_{_N}(k)}=-2\Im[\omega_{_N}(k)]=-2\sum_{m=2}^\infty W^{m}\Im[A_N^{(m)}(k)]
\end{equation}
so that $\mathrm{e}^{-t/\tau_{_N}(k)}=|\mathrm{e}^{-\mathrm{i}\omega_{_N}(k)t}|^2$.

\subsection*{Three-point structure factor}

Define
\begin{equation}
    \Xi_N(k_1,k_2)\equiv\frac{1}{N}\tilde{n}_{_N}(k_1)\tilde{n}_{_N}(k_2)\tilde{n}_{_N}(-k_1-k_2),
\end{equation}
whose thermodynamic limit is given by
\begin{equation}
    \lim_{N\to\infty}\expval{\Xi_N(k_1,k_2)}=S_3(k_1,k_2)+2\pi[S(k_1)\delta(k_2)+S(k_2)\delta(k_1)+S(k_1)\delta(k_1+k_2)]+(2\pi)^2\delta(k_1)\delta(k_2),
\end{equation}
where we define the three-point structure factor as
\begin{equation}
    S_3(k_1,k_2)\equiv1+\tilde{h}(k_1)+\tilde{h}(k_2)+\tilde{h}(-k_1-k_2)+\tilde{h}_3(k_1,k_2),
\end{equation}
where $\tilde{h}_3(k_1,k_2)$ is the Fourier transform of
\begin{equation}
    h_3(\Delta x_1,\Delta x_2)\equiv g_3(\Delta x_1,\Delta x_2)-g_2(\Delta x_1)-g_2(\Delta x_2)-g_2(\Delta x_1-\Delta x_2)+2.
\end{equation}

\subsection*{Evaluation of the high-loop propagators}

Based on the identity
\begin{equation}
    \frac{1}{x+\mathrm{i}0^+}\equiv\mathcal{P}\left(\frac{1}{x}\right)-\pi\mathrm{i}\delta(x),\hspace{2em}\frac{1}{(x+\mathrm{i}0^+)^2}\equiv-\left(\frac{1}{x+\mathrm{i}0^+}\right)'\equiv\mathcal{P}\left(\frac{1}{x^2}\right)+\pi\mathrm{i}\delta'(x),\ldots
\end{equation}
we have
\begin{equation}
    \frac{1}{(k^2-q^2+\mathrm{i}0^+)^m}=\mathcal{P}\left[\frac{1}{(k^2-q^2)^m}\right]-\frac{\pi\mathrm{i}}{(m-1)!}\delta^{(m-1)}(q^2-k^2).
\end{equation}
For example, for the first order and second order, we have
\begin{equation}
    \begin{cases}
        \begin{aligned}
        &\mathcal{P}\left(\frac{1}{k^2-q^2}\right)\to\frac{1-(\delta_{qk}+\delta_{q,-k})}{k^2-q^2},\\
        &\pi\mathrm{i}\delta(q^2-k^2)=\frac{\pi\mathrm{i}}{2k}[\delta(q-k)+\delta(q+k)]\to\frac{\mathrm{i}}{4k}N(\delta_{qk}+\delta_{q,-k}),
        \end{aligned}
    \end{cases}
\end{equation}
\begin{equation}
    \begin{cases}
        \begin{aligned}
            &\mathcal{P}\left[\frac{1}{(k^2-q^2)^2}\right]\to\frac{1-(\delta_{qk}+\delta_{q,-k})}{(k^2-q^2)^2}-\frac{1}{4k^2}\frac{\sum_{m=1}^\infty\frac{2}{m^2}}{(2\pi/N)^2}(\delta_{qk}+\delta_{q,-k})=\frac{1-(\delta_{qk}+\delta_{q,-k})}{(k^2-q^2)^2}-\frac{N^2}{48k^2}(\delta_{qk}+\delta_{q,-k}),\\
            &\pi\mathrm{i}\delta'(q^2-k^2)=\frac{\pi\mathrm{i}}{4k^2}[\delta'(q-k)+\delta'(q+k)]+\frac{\pi\mathrm{i}}{4k^3}[\delta(q-k)+\delta(q+k)]\to\frac{\mathrm{i}}{8k^2}N(\delta'_{qk}+\delta'_{q,-k})+\frac{\mathrm{i}}{8k^3}N(\delta_{qk}+\delta_{q,-k})
        \end{aligned}
    \end{cases}
\end{equation}
in the discrete limit, respectively.

\subsection*{Perturbative calculations and final results}

For the second order, there is no high-loop diagram term but only the 1-loop diagram shown in Fig.~\ref{fig:A2}, so we have
\begin{equation}
    A_N^{(2)}(k)=\mathcal{A}_N^{(2)}(k)=\frac{1}{N}\left(\frac{-\mathrm{i}}{4k}\right)\sum_{i,j=1}^N[2\mathrm{e}^{2\mathrm{i}k(x_{ij})_+}-1]=-\frac{\mathrm{i}}{4k}Z_N(2k)\implies A^{(2)}(k)=-\frac{\mathrm{i}}{4k}Z(2k).
\end{equation}
For the third order, noting that $\sum_{q_1,q_2\ne k}=\sum_{q_1\ne q_2\ne k}+\sum_{q_1=q_2\equiv q\ne k}$, we have
\begin{equation}
    \mathcal{A}_N^{(3)}(k)=A_N^{(3)}(k)+B_N^{(3)}(k)=-\frac{1}{16k^2}\frac{1}{N}\sum_{i,j,l=1}^N[2\mathrm{e}^{\mathrm{i}k(x_{ij})_+}-1][2\mathrm{e}^{\mathrm{i}k(x_{jl})_+}-1]=-\frac{J_N(2k)+1/3}{4k^2}+\frac{N^2}{48k^2},
\end{equation}
containing a divergent term $N^2/(48k^2)$ as we expected. The 2-loop diagram term [Fig.~\ref{fig:B3}] can be calculated as
\begin{equation}
    \begin{aligned}
        \mathcal{B}_N^{(3)}(k)=B_N^{(3)}(k)&=\frac{1}{N^3}\sum_{q\ne k}\frac{N|\tilde{n}_{_N}(k-q)|^2}{(k^2-q^2+\mathrm{i}0^+)^2}=\frac{N^2}{48k^2}+\frac{1}{N}\sum_q\mathcal{P}\left[\frac{\mathscr{S}_N(k-q)}{(k^2-q^2)^2}\right]-\frac{\mathrm{i}}{8k^3}[k\mathscr{S}_N'(2k)+\mathscr{S}_N(2k)],
    \end{aligned}
\end{equation}
and $A_N^{(3)}(k)$ [Fig.~\ref{fig:A3}] is given by
\begin{equation}
    \begin{aligned}
        A_N^{(3)}(k)&=\mathcal{A}_N^{(3)}(k)-\mathcal{B}_N^{(3)}(k)=-\frac{J_N(2k)+1/3}{4k^2}-\text{f.p.}\int_{-\infty}^\infty\frac{\mathrm{d}q}{2\pi}\left[\frac{\mathscr{S}_N(k-q)}{(k^2-q^2)^2}\right]+\frac{\mathrm{i}}{8k^3}[k\mathscr{S}_N'(2k)+\mathscr{S}_N(2k)],
    \end{aligned}
\end{equation}
whose thermodynamic limit becomes
\begin{equation}
    A^{(3)}(k)\equiv\lim_{N\to\infty}\expval{A_N^{(3)}(k)}=-\frac{K(2k)+1/3}{4k^2}-\text{f.p.}\int_{-\infty}^\infty\frac{\mathrm{d}q}{2\pi}\frac{ S(k-q)}{(k^2-q^2)^2}+\frac{\mathrm{i}}{8k^3}[kS'(2k)+S(2k)],
\end{equation}
where $\text{f.p.}\int f(q)\mathrm{d}q\equiv\int\mathcal{P}[f(q)]\mathrm{d}q$ denotes the Hadamard finite part of the integral with high-order singularities, which can be viewed as a generalization of the principal value.
For the fourth order, $A_N^{(4)}(k)$ [Fig.~\ref{fig:A4}] can be proved to be given by
\begin{equation}
    \begin{aligned}
        &A_N^{(4)}(k)=\mathcal{A}_N^{(4)}(k)-\mathcal{B}_N^{(4a)}(k)-\mathcal{B}_N^{(4b)}(k)-\mathcal{B}_N^{(4c)}(k)+2\mathcal{C}_N^{(4)}(k)\\
        ={}&\frac{\mathrm{i}}{32k^3}\left[4H_N(2k)+2Q_N(2k)+Z_N(4k)\right]-\frac{\mathrm{i}}{16k^5}[k^2\mathscr{S}_N''(2k)+2k\mathscr{S}_N'(2k)+\mathscr{S}_N(2k)]\\
        &+\frac{\mathrm{i}}{2k}\times\text{f.p.}\int_{-\infty}^\infty\frac{\mathrm{d}q}{2\pi}\frac{\Re[\Xi_N(q-k,2k)]}{(k^2-q^2)^2}+\frac{\mathrm{i}}{4k}\mathscr{S}_N(2k)\times\text{f.p.}\int_{-\infty}^\infty\frac{\mathrm{d}q}{2\pi}\frac{\mathscr{S}_N(k-q)}{(k^2-q^2)^2}\\
        &+\frac{\mathrm{i}}{4k^3}\times\text{p.v.}\int_{-\infty}^\infty\frac{\mathrm{d}p}{2\pi}\frac{\Re[\Xi_N(p-k,2k)+k\partial_2\Xi_N(p-k,2k)]}{k^2-p^2}\\
        &+\frac{\mathrm{i}}{8k^3}[\mathscr{S}_N(2k)+k\mathscr{S}_N'(2k)]\times\text{p.v.}\int_{-\infty}^\infty\frac{\mathrm{d}p}{2\pi}\frac{\mathscr{S}_N(p+k)}{k^2-p^2}+\text{f.p.}\int_{-\infty}^\infty\frac{\mathrm{d}q}{2\pi}\frac{2\mathscr{S}_N(k-q)}{(k^2-q^2)^3}\\
        &-\text{f.p.}\int_{-\infty}^\infty\int_{-\infty}^\infty\frac{\mathrm{d}q\mathrm{d}p}{(2\pi)^2}\frac{2 \Re[\Xi_N(k-q,p-k)]}{(k^2-q^2)^2(k^2-p^2)}-\text{f.p.}\int_{-\infty}^\infty\int_{-\infty}^\infty\frac{\mathrm{d}q\mathrm{d}p}{(2\pi)^2}\frac{\mathscr{S}_N(k-q)\mathscr{S}_N(q-p)}{(k^2-q^2)^2(k^2-p^2)},
    \end{aligned}
\end{equation}
with its thermodynamic limit being
\begin{equation}
    \begin{aligned}
        &A^{(4)}(k)=\lim_{N\to\infty} \expval{A_N^{(4)}(k)}\\
        ={}&\frac{\mathrm{i}}{32k^3}\left[4\Omega(2k)+2Q(2k)+Z(4k)-4Z''(2k)\right]-\frac{\mathrm{i}}{16k^5}[k^2S''(2k)+2kS'(2k)+S(2k)]\\
        &+\frac{\mathrm{i}}{2k}\times\text{f.p.}\int_{-\infty}^\infty\frac{\mathrm{d}q}{2\pi}\frac{\Re[S_3(q-k,2k)]}{(k^2-q^2)^2}+\frac{\mathrm{i}}{4k}S(2k)\times\text{f.p.}\int_{-\infty}^\infty\frac{\mathrm{d}q}{2\pi}\frac{S(k-q)}{(k^2-q^2)^2}\\
        &+\frac{\mathrm{i}}{4k^3}\times\text{p.v.}\int_{-\infty}^\infty\frac{\mathrm{d}p}{2\pi}\frac{\Re[S_3(p-k,2k)+k\partial_2S_3(p-k,2k)]}{k^2-p^2}\\
        &+\frac{\mathrm{i}}{8k^3}[S(2k)+kS'(2k)]\times\text{p.v.}\int_{-\infty}^\infty\frac{\mathrm{d}p}{2\pi}\frac{S(p+k)}{k^2-p^2}-\text{f.p.}\int_{-\infty}^\infty\frac{\mathrm{d}q}{2\pi}\frac{S(k-q)}{(k^2-q^2)^3}\\
        &-\text{f.p.}\int_{-\infty}^\infty\int_{-\infty}^\infty\frac{\mathrm{d}q\mathrm{d}p}{(2\pi)^2}\frac{2 \Re[S_3(k-q,p-k)]}{(k^2-q^2)^2(k^2-p^2)}-\text{f.p.}\int_{-\infty}^\infty\int_{-\infty}^\infty\frac{\mathrm{d}q\mathrm{d}p}{(2\pi)^2}\frac{ S(k-q)S(q-p)}{(k^2-q^2)^2(k^2-p^2)}.
    \end{aligned}
\end{equation}

The inverse of the lifetime $1/\tau_{_N}(k)$ up to the order of $W^4$ is given by
\begin{equation}\label{eq:inversetauN}
    \frac{1}{\tau_{_N}(k)}=\frac{W^2}{2k}\mathscr{S}_N(2k)+\frac{W^3}{4k^2}\left[2\Im[J_N(2k)]-\mathscr{S}_N'(2k)-\frac{1}{k}\mathscr{S}_N(2k)\right]-2 W^4\Re[A_N^{(4)}(k)]+O(W^5),
\end{equation}
whose thermodynamic limit is
\begin{equation}
    \frac{1}{\tau_k}\equiv\lim_{N\to\infty}\expval{\frac{1}{\tau_{_N}(k)}}=\frac{W^2}{2k}S(2k)-\frac{W^3}{4k^3}\left[3kS'(2k)+S(2k)\right]-2 W^4\Re[A^{(4)}(k)]+O(W^5).
\end{equation}
For disordered stealthy hyperuniform systems, $S(k)\equiv0$ for $0\le k<K=2\pi\chi$, and we have
\begin{equation}
    \frac{1}{\tau_k}=-\frac{ W^4}{16k^3}\Re[4\Omega(2k)+2Q(2k)+S(4k)]+O(W^5)\approx\frac{ W^4}{16k^3}[4c-S(4k)]+O(W^5)
\end{equation}
for $0<k<K/2=\pi\chi$, where $c$ is given by Eq.~\eqref{eq:c}.
In this sense, we conclude that the stealthy hyperuniform system is transparent up to order $W^3$ in the transparent interval $0<k<\pi\chi$, consistent with the free-boundary-condition conclusion.
Besides, plugging $v_{_N}(k)=2k+O(W)$ into Eq.~\eqref{eq:inversetauN}, the corresponding Lyapunov exponent $\lambda_{_N}(k)$ can be calculated as
\begin{equation}
    \lambda_{_N}^{\text{pbc}}(k)=\frac{1}{v_{_N}(k)\tau_{_N}(k)}=\frac{W^2}{4k^2} \mathscr{S}_N(2k)+O(W^3),
\end{equation}
also agreeing with the free-boundary-condition result $\lambda_{_N}^{\text{fbc}}(k)$.

\end{document}